\documentclass[12pt,preprint]{aastex}

\newcommand{\etal}{{et al.\ }}

\begin{document}

\newcommand{\gsim}{\raisebox{-.4ex}{$\stackrel{>}{\scriptstyle \sim}$}}
\newcommand{\lsim}{\raisebox{-.4ex}{$\stackrel{<}{\scriptstyle \sim}$}}
\newcommand{\psim}{\raisebox{-.4ex}{$\stackrel{\propto}{\scriptstyle \sim}$}}
\newcommand{\kms}{\mbox{km~s$^{-1}$}}
\newcommand{\s}{\mbox{$''$}}
\newcommand{\mloss}{\mbox{$\dot{M}$}}
\newcommand{\my}{\mbox{$M_{\odot}$~yr$^{-1}$}}
\newcommand{\ls}{\mbox{$L_{\odot}$}}
\newcommand{\ms}{\mbox{$M_{\odot}$}}
\newcommand{\mm}{\mbox{$\mu$m}}
\newcommand{\water}{\mbox{H$_2$O}}
\newcommand{\methanol}{\mbox{CH$_3$OH}}
\def\arcdeg{\hbox{$^\circ$}}
\newcommand{\secp}{\mbox{\rlap{.}$''$}}%

\shortauthors{Rioja, et al.}
\shorttitle{Astrometry at 130\,GHz with the KVN}

\title{The Power of Simultaneous Multi-Frequency Observations for mm-VLBI: Astrometry up to 130\,GHz with the KVN.}

 \author{
   Mar\'{\i}a J. \textsc{Rioja}\altaffilmark{1,2},
   Richard \textsc{Dodson}\altaffilmark{1},
   Taehyun \textsc{Jung}\altaffilmark{3,4},
   Bong Won \textsc{Sohn}\altaffilmark{3,4}
}
 \affil{$^1$ International Centre for Radio Astronomy Research, M468,
The University of Western Australia, 35 Stirling Hwy, Crawley, Western Australia, 6009}
 \affil{$^2$ Observatorio Astron\'omico Nacional (IGN), Alfonso XII, 3 y 5, 28014 Madrid, Spain} 
 \affil{$^3$ Korea Astronomy and Space Science Institute, Daedeokdae-ro 776, Yuseong-gu, Daejeon 305-348, Korea}
 \affil{$^4$ University of Science and Technology, 217 Gajeong-ro Yuseong-gu, Daejeon, 305-350, Korea}
 \email{maria.rioja@icrar.org}


\begin{abstract} 

Simultaneous observations at multiple frequency bands have the potential to overcome the fundamental limitation imposed by the atmospheric propagation in mm-VLBI observations. 
The propagation effects place a severe limit in the  sensitivity achievable in mm-VLBI, reducing 
the time over which the signals can be coherently combined, and preventing the use of phase referencing and astrometric measurements.
We carried out simultaneous observations at 22, 43, 87 and 130\,GHz of a group of five AGNs, the weakest of which is $\sim$200\,mJy at 130\,GHz, with angular separations ranging from 3.6 to 11 degrees, using the KVN. 
We analysed this data using the Frequency Phase Transfer (FPT) and the Source Frequency Phase Referencing (SFPR) techniques, which use the observations at a lower frequency to correct those  at a higher frequency.
The results of the analysis provide an empirical demonstration of the increase in the coherence times at 130\,GHz from a few tens of seconds to about twenty minutes, with FPT, and up to many hours with SFPR. 
Moreover the astrometric analysis provides high precision relative position measurements between two frequencies, including, for the first time, astrometry at 130\,GHz.
Finally we demonstrate a method for the generalised decomposition of the relative position measurements into absolute position shifts for bona fide astrometric registration of the maps of the individual sources at multiple frequencies, up to 130\,GHz.

\end{abstract}

\section{Introduction}\label{sec:intr}

Astronomical studies by means of Very Long Baseline Interferometry (VLBI) observations at cm wavelengths is a well established field, with advanced technological developments and analysis techniques 
that result in superb quality images, including those of very weak  $\mu$Jy sources (e.g. \citet{garret_micro}) and with micro-arcsecond ($\mu$as) astrometry measurements (e.g. \citet{reid_micro}), using phase referencing techniques. This is applied to a wide variety of targets and fields of study.

VLBI at (sub)mm wavelengths ({\it hereafter} mm-VLBI) can result in the highest angular resolutions achieved in astronomy and has a unique access to emission regions that are inaccessible with any other approach or at longer wavelengths, because the compact areas of interest are often self-absorbed.
Therefore it holds the potential to increase our understanding of the physical processes in e.g. Active Galactic Nuclei (AGN) and in the vicinity of supermassive Black Holes, and for studies of molecular transitions at high frequencies. 

Nevertheless the applications of mm-VLBI are much less widespread.
The observations become progressively more challenging as the wavelength gets shorter because of the: limited telescope surface accuracy and aperture efficiency, receiver system temperatures and sensitivity, shorter atmospheric coherence times and  that sources are intrinsically weaker in general. Moreover phase referencing techniques,which are routinely used in cm-VLBI, fail to work beyond 43\,GHz 
(excluding a single case at 86\,GHz with the Very Long Baseline Array (VLBA) \citep{porcas_02}).

Continuous development and technical improvements have led to a sustained 
increase of the high frequency threshold for VLBI observations in the last two decades (see \citet{krichbaum_mmvlbi} for a review). Regular observations up to 86\,GHz are being carried out 
with well established networks such as the VLBA and the Global Millimeter VLBI Array (GMVA), more recently up to 130\,GHz with the Korean VLBI Network (KVN) and ad-hoc observations at the highest frequencies up to 240\,GHz \citep{sgra_240}. 
The field of mm-VLBI will greatly benefit from the arrival of the Phased-up Atacama Large Millimeter Array (ALMA) \citep{alma_pp} for joint VLBI observations.

In this paper we will focus on two aspects that limit the potential of mm-VLBI observations: 1) achieving improved sensitivity through increased coherence times, to increase the number of targets; 2) achieving astrometry, and in particular for ``bona fide''  registration of images at multiple frequency bands, to reveal the physical processes in a number of fields of astronomy.  

For example, for AGN studies, maps of the Spectral Index or Rotation Measure across the source, at mm-wavelengths, provide crucial insights into the development of the magnetic field strength and particle densities as the jet exits the core region and extends down the out-flow.
%
However, astrometric map registration is crucial to make a reliable measurement and to form meaningful interpretation. There are a number of methods in AGN studies which can be used to align images at multiple frequencies, as discussed in \citet{hovatta_14}. They  argued that any results derived without accurate astrometric registration are questionable in the vicinity of the core, which is the most interesting region in mm-VLBI. 
Also, in studies of the maser emission from the molecular species that exist in circumstellar envelopes (CSEs) and star forming regions (SFR), the comparison of the locations 
of the different species of maser emission can be inverted to reveal the physical conditions as a function of the distance from the central star pumping the masers (e.g. see discussion in \citet{reid_moran_masers}). In both cases this would allow one to fully understand the flow of material and energy in stellar environments during the formation and the evolution of stars. 
Traditionally 
there has been no other mechanism other than phase referencing to accurately astrometrically register the maps at the different bands. 

The Korean VLBI Network (KVN) \citep{early_kvn,sslee_14} is the first dedicated mm-VLBI array and addresses one of the fundamental limitations of the field, the atmospheric stability. It currently consists of three antennas operated by Korea Astronomy and Space Science Institute (KASI), spread across South Korea, located in the campus of the Universities of Yonsei and Ulsan in main land and on Jeju island. The observing frequencies are  centred at 22, 43, 87 and 130\,GHz. 
The baseline lengths between the antennas range between 300 and 500\,km, which provide a spatial resolution $\sim$1\,mas at the highest frequency band.
The innovative multi-band receiver \citep{han_08,kvn_optics} of KVN is designed to mitigate the atmospheric propagation effects 
using simultaneous observations at multiple bands.
%
The KVN combined with the Frequency Phase Transfer (FPT) and Source Frequency Phase Referencing (SFPR) data analysis techniques \citep{vlba_31,rioja_11a,rioja_11b, rioja_14} (see also references therein) allows an effective increase of the coherence time,  well beyond that imposed by tropospheric fluctuations, as well as high precision astrometric measurements, respectively,  even at the highest frequencies.
We know of no demonstrated upper frequency limit and the methods would be expected to work as long as the tropospheric propagation effects were non-dispersive. \\
In this context, successful tests are on-going with ALMA \citep{alma_test} at frequencies as high as 650\,GHz (where this is known as the Band-to-Band (B2B) mode).
In \citet{rioja_14} we presented results of SFPR astrometric measurements with KVN at 22 and 43\,GHz for continuum sources, along with a detailed comparative study using fast frequency switching  observations with the VLBA. \citet{dodson_14} presented the application of SFPR to spectral line studies for astrometric registration of the H$_2$O and SiO maser maps, at 22 and 43\,GHz, respectively, in CSEs. In this paper we extend the astrometric measurements to all the four frequency bands supported by the KVN, up to 130\,GHz, and quantify the increase in coherence time. 
The paper layout is as follows: The simultaneous multi-frequency observations at four-bands are presented in Section 2; a description of the analysis carried out to obtain the maps,
astrometric measurements and ``bona fide'' astrometric registration of multi-frequency images is in Section 3; the results are presented in Section 4 and a discussion of the results in Section 5.

\section{Observations}\label{sec:obs}

In March 5, 2014, we carried out simultaneous observations at four frequencies, i.e. 22, 43, 87 and 130\,GHz (also known as K, Q, W and D bands, respectively), with the three antennas of the KVN, towards 5 AGN target sources, for a total duration of 9 hours. 

The recording consisted of 4 consecutive 16 MHz intermediate frequency (IF) sub-bands at each frequency. The lower edges of the first IFs at Q, W and D bands were selected to be multiples of that at K band, those being 21.65, 43.30, 86.60 and 129.90\,GHz.
Having integer frequency ratios is important for the successful application of tropospheric compensation techniques using multi-frequency observations.
The correlation was done with the DiFX correlator \citep{difx} with 1 second averaging and a spectral resolution of 64 channels per IF. 

\begin{figure}
\centering
\includegraphics[width=0.95\textwidth]{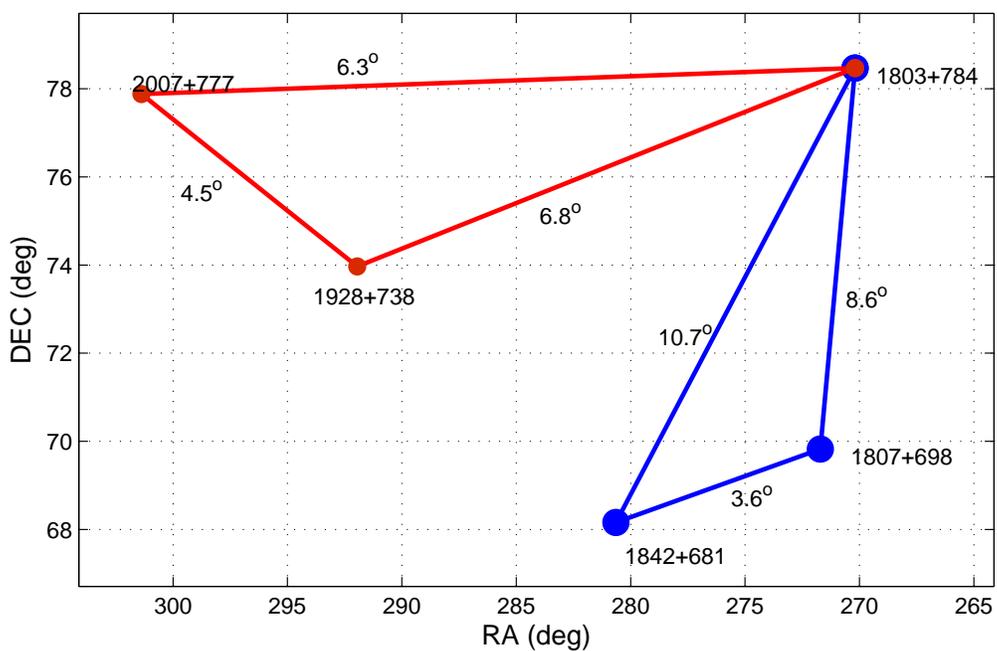}
\caption{Sky distribution of the five AGN sources in this study. The source pairs in the SFPR analysis are connected with a line, and  have angular separations ranging 
from 3.6$^o$ to 11$^o$. The triangles connect sources that were observed in a $\sim$1.5 hour block,  alternating between the two triangles.
}
\label{fig:source_pos}
\end{figure}

The target sources were selected from the 86-GHz VLBI catalog \citep{sslee_86} based on two criteria: to have strong detections at W band, and angular separations in the sky ranging from a few to many degrees. 
Figure \ref{fig:source_pos} shows the distribution in the sky of the five selected sources (1803+784, 1807+698, 1842+681, 1928+738 and 2007+777) 
along with their angular separations, which range between $\sim 3.6^o$  and $11^o$. 
None of the sources had been observed previously with VLBI at 130\,GHz.
The observations consisted of $\sim$3 minutes long scans alternating between the sources in each of the two triangles shown in Figure \ref{fig:source_pos}, and between the triangles ca. every hour and a half, at the four bands simultaneously. Alternating between multiple sources allowed us to develop a strategy to decompose the relative astrometric measurements into single source position shifts that allow, for example, the registration of the images at different frequencies.
%

\section{Methods}\label{sec:meth}

In this section we describe the various mapping and astrometric analyses carried out: 

\subsection{Hybrid Maps at the Four KVN Frequency Bands.}
We followed standard procedures for imaging VLBI datasets using {\sc AIPS} \citep{aips_cook}. 
One of the major challenges of imaging KVN observations arises from the small number of antennas, which prevents the application of amplitude self-calibration techniques to derive amplitude gain corrections, as this requires a minimum of four antennas. 

\begin{figure}
\centering
\includegraphics[width=0.95\textwidth]{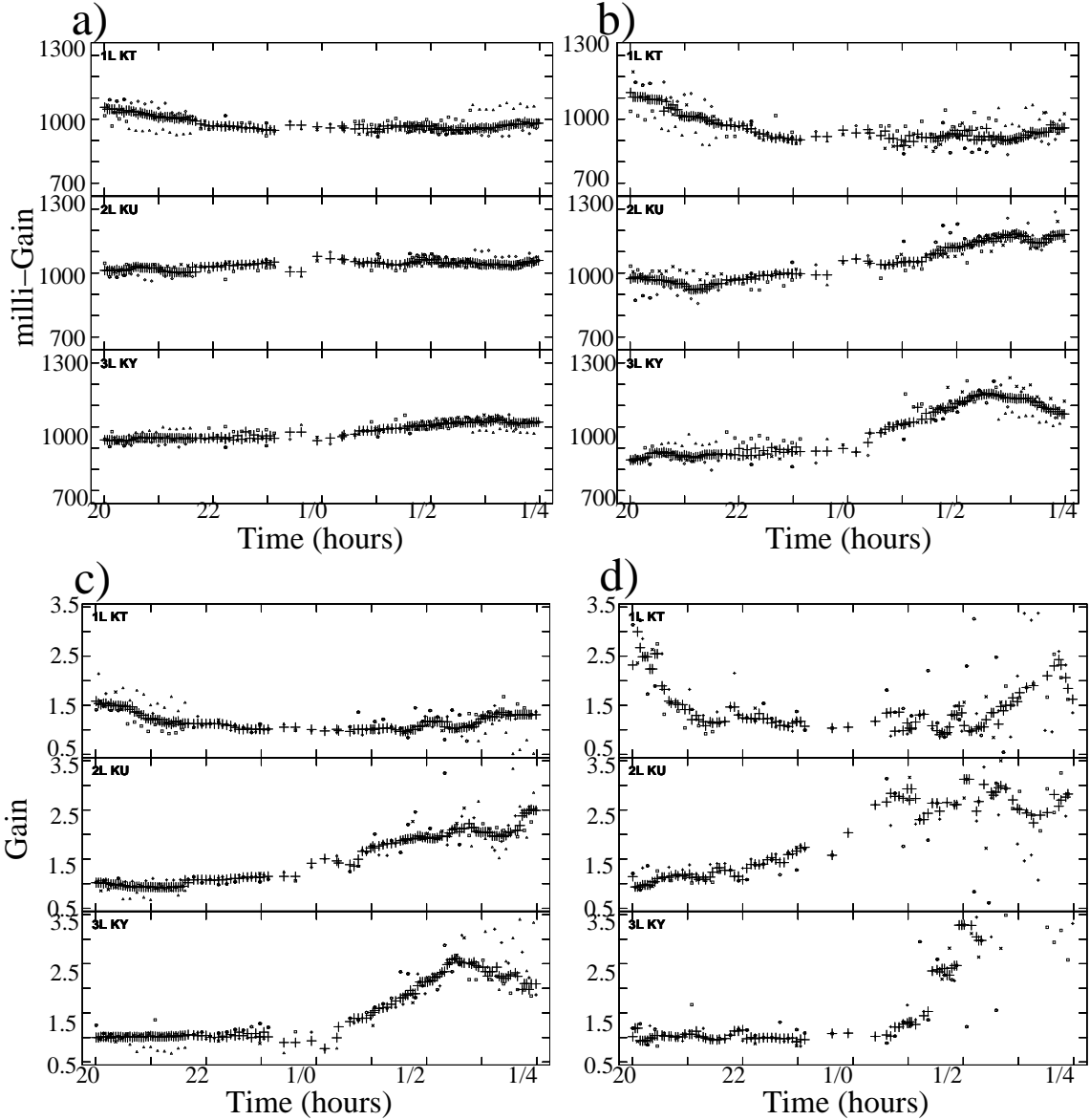}
\caption{Superimposed normalized amplitude gain correction factors to the nominal KVN calibration, versus time,
derived assuming point source models for each source (small symbols: `x', circles, squares, triangles and diamonds for 1803+784, 1807+698, 1842+681, 1928+738 and 2007+777 respectively), at all bands: 
a) 22\,GHz, b) 43\,GHz, c) 87\,GHz and d) 130\,GHz. At each band, the corrections from the individual sources were merged and smoothed to provide the final set of amplitude corrections, which were globally applied (large black crosses). At the lower frequencies (i.e. 22\,GHz and 43\,GHz) the estimated corrections were close to unity. On the other hand, at the higher frequencies  the corrections spanned a greater range both above and below the nominal value, with long periods where the gain corrections are less than unity. That implies significantly better than 100\% efficiency, which is unlikely, therefore these gains were adjusted so that the maximum efficiency correction was 100\%. These occurred at the highest elevations where one would expect the least deviation from the nominal gain performance.}
\label{fig:amp_sols}

\end{figure}

At the lower frequencies (K and Q bands), the system temperature measurements, along with regular sky-dipping, have been shown to provide a good estimate of the system performance \citep{sslee_14}. However, at the  higher frequencies (W and D bands) significant discrepancies can be expected. Hence, we have attempted to use the observations of the five target sources to estimate global (i.e. for all sources) amplitude gain correction factors that should rescale the nominal calibration information, at each band.
We assumed a point source model of arbitrary flux (for all sources) and calculated the normalized amplitude gain corrections, for each individual source and at each frequency band.
The individual gain estimates for all sources at a given band showed a good agreement, as shown in 
Figure \ref{fig:amp_sols}; this supports the validity of the assumption of point source structure with KVN resolutions at all bands.
At each band, the values for all sources were merged and smoothed together, thereby further suppressing any individual source structure contributions to the estimated gain amplitudes,  except for one of the weakest sources, 2007+777, at the highest frequency, 130GHz, which has noisy solutions. The resultant amplitude gain correction factors were applied to the corresponding datasets using {\sc AIPS}.  
The hybrid maps, made with  {\sc difmap} \citep{difmap}, are presented in Section \ref{sec:res}.

\subsection{SFPR Maps and Astrometric Analysis}

We carried out SFPR astrometric analysis of the four-band multi-frequency KVN dataset using {\sc AIPS}. 
The details of SFPR analysis are presented elsewhere \citep{vlba_31,rioja_11a,rioja_11b,rioja_14,dodson_14} and basically consists of two calibration steps.
In a first step, the observations at the higher frequency bands ($\nu^{\rm high}$) are calibrated using the simultaneous observations at a lower frequency band ($\nu^{\rm low}$), for each source. This is done for all frequency pairs which have an integer frequency ratio R (with R = $\nu^{\rm high} \over \nu^{\rm low}$), which in turn is used to scale up the phase-calibration solutions at the lower frequency. 
This dual frequency  calibration step eliminates the common non-dispersive residual errors (e.g. tropospheric propagation effects and inaccurate coordinates) in the complex visibilitity output of the correlator, providing an increased signal coherence at the higher frequency. We have dubbed this step ``Frequency Phase Transfer'' (FPT) and the outcome FPT-ed visibilities. 
The second step of the calibration removes the remaining dispersive residual errors (i.e. instrumental and ionospheric propagation effects) using the interleaving observations of another source. This two-step calibration  retains the astrometric signature of any source position shifts between the two frequencies in the interferometric phase.  
The resultant calibrated visibilities, for a given frequency pair and a source pair, are dubbed SFPR-ed visibilities. 
The Fourier transformation of the SFPR dataset is the SFPR map, which conveys a bona fide astrometric measurement of the relative separation or shift between the position of the reference points in the images at the two frequencies,  for the two sources. 
If the reference points are selected to be the ``core'' components this shift corresponds to the ``core-shift'' phenomenae, or change of position at the base of the jet, due to frequency dependent opacity effects as defined by \citet{bk_79,lob_98}, for AGN observations. 
In general, the KVN resolution will result in structure blending of the ``core'' and other jet components into a single unresolved component in the maps, therefore this shift gives the relative angular separation between the centroids of the brightness distributions at the two frequencies, for the two sources.
Through-out this text we are using the term ``position-shift'' 
(rather than ``core-shift'') to include all cases. 
For a detailed description of structure blending effects in astrometric analysis see Rioja et al. 2014.

In this paper, 
the SFPR analysis comprises of five pairs of frequency bands, with values of R ranging from 2 to 6, and six pairs of sources, with angular separations between 3.6$^o$ and 11$^o$. The frequency band pairs along with the corresponding scaling factors for the lower frequency (R) in parenthesis are: K$\rightarrow$Q ($\times$2), K$\rightarrow$W ($\times$4), K$\rightarrow$D($\times$6), Q$\rightarrow$W($\times$2) and Q$\rightarrow$D($\times$3). The six pairs of sources are shown in Figure  \ref{fig:source_pos}  with connecting lines. 

\subsection{Astrometric Registration of Images Across Frequency Bands} 

The outcome of the SFPR  analysis are relative astrometric measurements and, just like the measurements from  phase referencing analysis, there is an inherent ambiguity as to what are the individual contributions arising from each of the two sources; that is, the solutions are degenerate.
Increasing the number of sources sets stronger constraints in the disentangling into individual contributions, as the additional constraints reduces, 
but does not break the degeneracy. Our observations comprise of multiple pairs of sources for this purpose. Starting with the pairwise  astrometric SFPR measurements,  we 
estimated the single source position-shifts, using the Singular Value Decomposition (SVD) as the linear least squares minimization method, for each frequency pair. 

No matter how many combinations of sources one measures there will always remain an ambiguity of the global (i.e. common for all sources) absolute correction.  That is, for example, in the case that all the sources have identical position-shifts it would leave no signature in the measurements. 
Therefore, once we have decomposed the measurements into the contributions from the individual sources, we still need to find the global absolute correction. 
We have additional information which allows us to do this in most cases, namely, that a frequency dependent source position shift 
is expected to be aligned with the direction of the jet in the map of the source.
This expectation applies for both types of position shifts described above, i.e. core-shifts arising from opacity effects and/or 
centroid shifts arising from structure blending, regardless of its nature. 
This approach will fail for the cases when the position-shifts and the jet axis do not coincide, which would be unexpected, or when all the sources have similar jet directions and therefore there is no clear best global correction to be determined.
The group of sources in our observations shows a wide range of jet directions in the high resolution VLBI MOJAVE maps \citep{mojave}. 
Hence, by adding the constraint that the position-shift direction must align with the up-stream jet direction we can determine both the appropriate global correction and unambiguous individual source position-shifts (also called absolute position shifts hereafter). The latter are also the shifts required for a bona-fide astrometric registration of the images at the four observed frequencies, for all the sources. 

\section{Results}\label{sec:res}

\subsection{Hybrid Maps at the Four KVN Frequency Bands.} 

\begin{figure}
\centering
\includegraphics[width=0.9\textwidth]{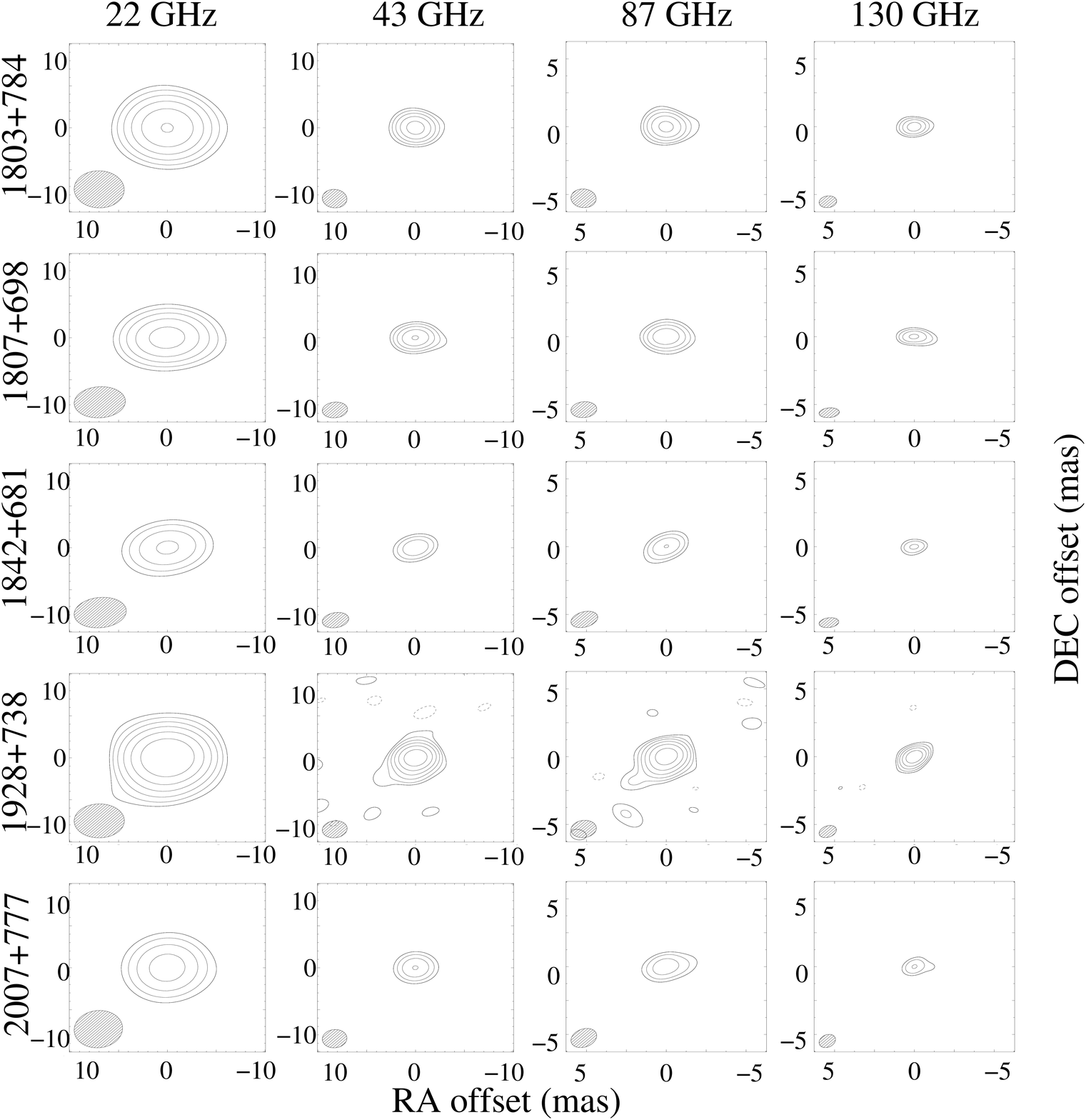}
\caption{Self-calibrated images of the five AGNs at the four KVN frequency bands. The contours in all cases start at $\pm$50mJy and double for each subsequent contour line. The image sizes are $\pm$10 mas at K and Q bands, and $\pm$5 mas at W and D bands. The beam size is indicated by the elipse at the bottom left of the image.}
\label{fig:sc}
\end{figure}

Figure \ref{fig:sc} shows self-calibration images for the five AGN sources in the multi-frequency KVN observations, including the first images of these sources at 130\,GHz. 
The visibility datasets were modelfitted and imaged 
 and show little divergence from point-source core dominated images; in some cases there appears to be some elongation aligned with the jet direction as seen in higher resolution maps from MOJAVE \citep{mojave}. 

%
Table \ref{tab:flux} lists the total flux values  as measured from the maps, at all frequency bands. It should be noted that the absolute flux values might suffer from 
the lack of absolute calibration, especially at the highest frequency band, 130\,GHz (see section 3.1 for a description of the amplitude calibration). 

It is worth emphasizing that not all sources had direct detections, i.e. within the atmospheric coherence time, 
and for those that didn't, we benefited from the extended coherence time resulting from a previous trans-frequency FPT analysis calibration. 
Note that remaining dispersive residuals prevent making a map after solely FPT calibration; nevertheless the FPT analysis conditions the dataset at $\nu^{\rm high}$  and allows 
for a self-calibration analysis using much longer phase solution time intervals, hence enabling the detection of sources that would not be detected otherwise 
(i.e. within the atmospheric coherence time interval). The resulting maps are therefore self-calibration maps, and have no astrometry information. 
This procedure enabled the imaging of 2007+777 and 1842+681 at 130\,GHz, which were too weak for direct detections.

\begin{table}[h]
\centering
\begin{tabular}{l|cccc}
\hline \hline
        & 22\,GHz & 43\,GHz & 87\,GHz & 130\,GHz \\
\hline
1803+784 & 1.7 & 1.2 & 1.0 & 0.6 \\
1807+698 & 1.1 & 0.9 & 0.9 & 0.5 \\
1842+681 & 0.5 & 0.4 & 0.4 & 0.2 \\
1928+738 & 3.7 & 3.2 & 3.3 & 1.6 \\
2007+777 & 0.6 & 0.4 & 0.4 & 0.3\\
\hline
\end{tabular}
\caption{Total Source flux, in Jy, 
for the five sources and at the four frequency bands, measured from the self-calibrated maps in Figure \ref{fig:sc}. 
\label{tab:flux}}
\end{table}

\subsection{Increased Coherence Time for mm-VLBI} 

\begin{figure}
\centering
 \includegraphics[width=\textwidth]{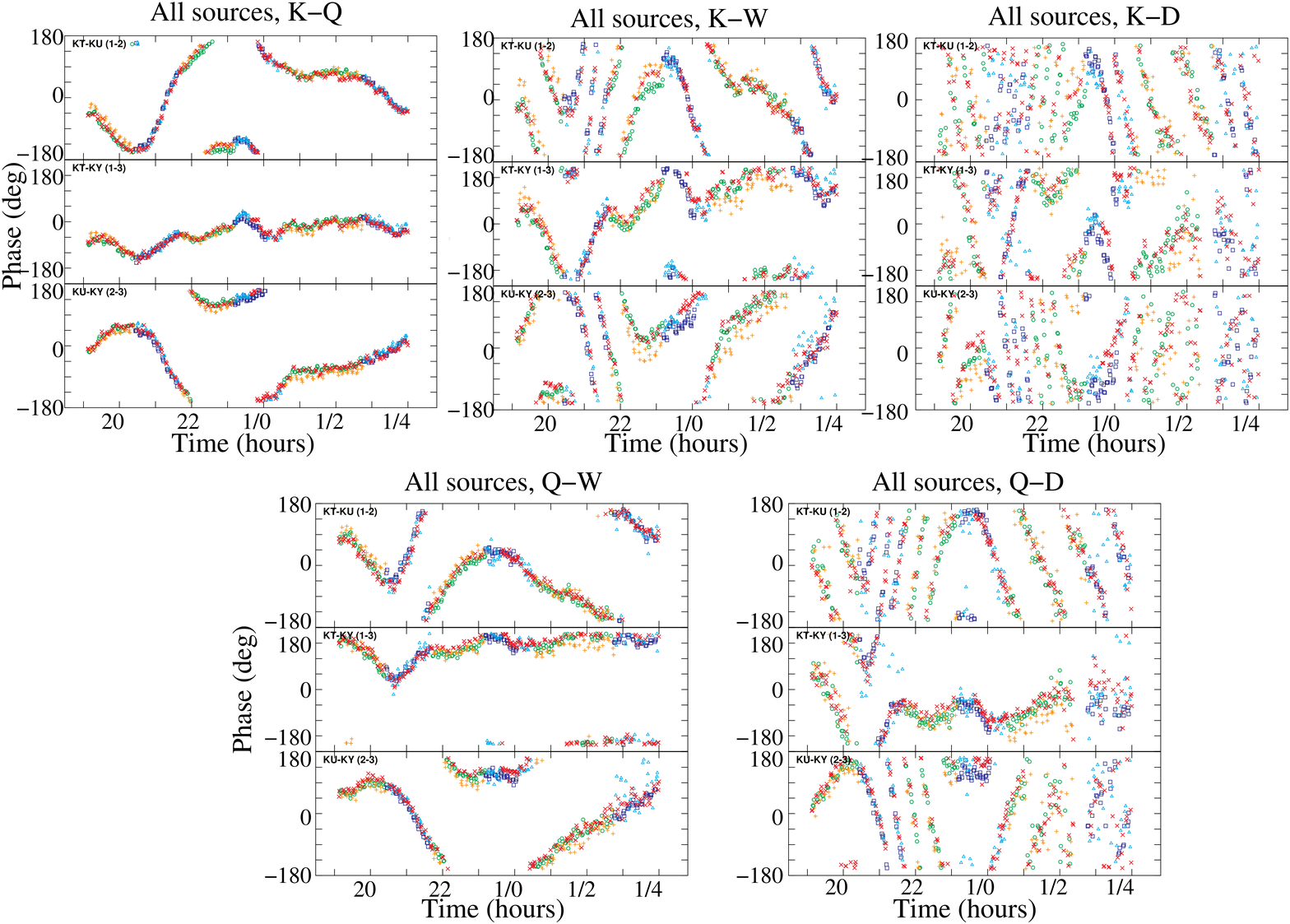}
\caption{Superimposed FPT-ed residual visibility phases for all the sources in this study, in separate plots for each of the five frequency pairs ($\nu^{\rm low} \rightarrow \nu^{\rm high}$) that have 
an integer frequency ratio R. Note that the FPTed visibility phases are the correlator output at $\nu^{\rm high}$ solely calibrated with the scaled up phase solutions at $\nu^{\rm low}$. 
 {\it Upper row}: Corresponds to frequency pairs with $\nu^{\rm low} \,=\,22$\,GHz (K band). From left to right,  K$\rightarrow$Q (R=2),  K$\rightarrow$W (R=4) and  K$\rightarrow$D (R=6). {\it Lower row}: Corresponds to frequency pairs with  $\nu^{low}\,=\,43$\,GHz (Q band). From left to right,  Q$\rightarrow$W (R=2) and Q$\rightarrow$D (R=3). 
Different colours and symbols correspond to observations of different sources: red (`x') for 1803+784, green (`circle') for 1807+698, orange (`+') for 1842+681, dark blue (`square') for 1928+738 and cyan (`triangle') for 2007+777. The enhanced coherence time is immediately obvious as is the agreement between the phases for the different sources, for each frequency pair. The rightmost plots underline the benefits from using the scaled calibration values derived from $\nu^{\rm low} \,=\,43$\,GHz calibration for the 130-GHz data, as compared to $\nu^{\rm low} \,=\,22$\,GHz.}
\label{fig:fpt_vis}
\end{figure}

\begin{figure}
\centering
\includegraphics[width=1.2\textwidth]{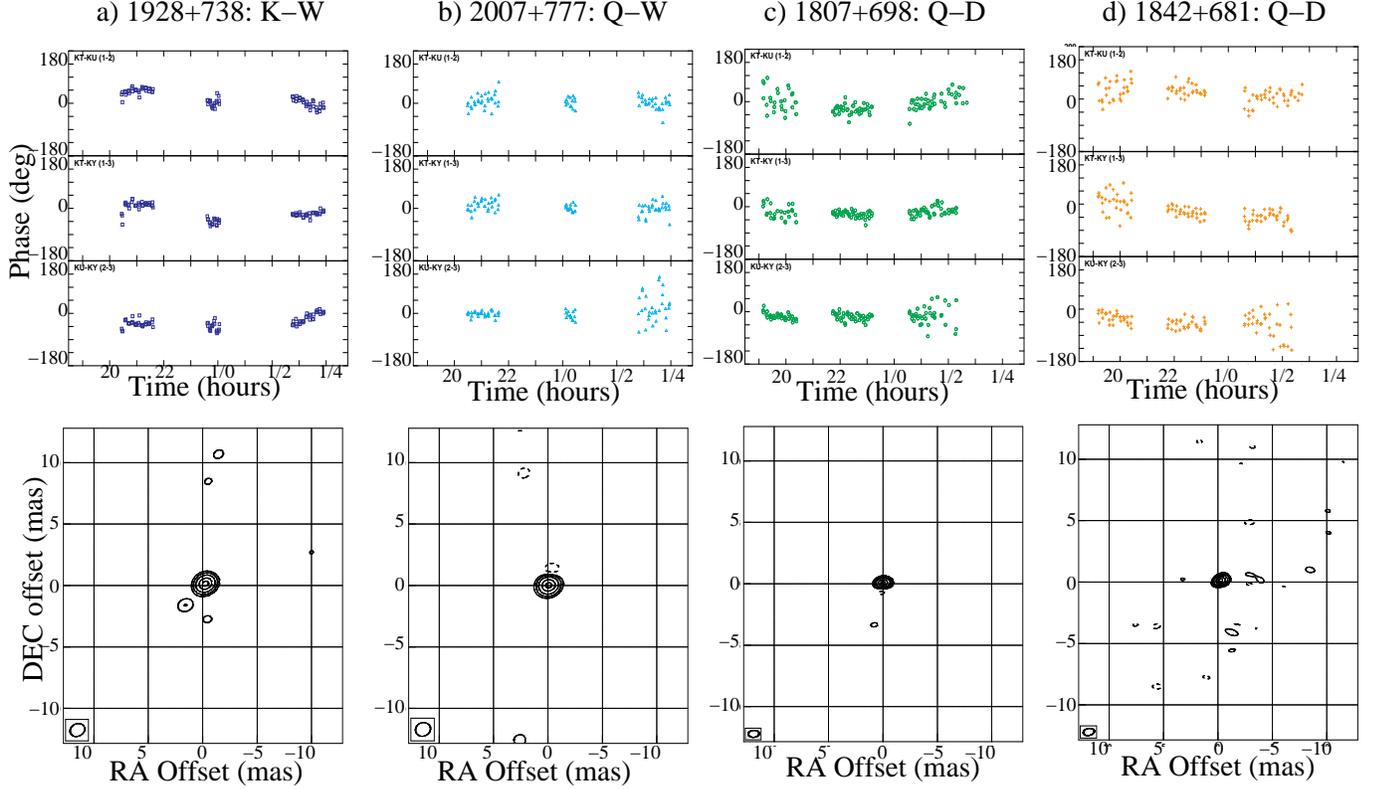}
\caption{Outcomes of SFPR astrometric analysis: {\it Upper Row}: SFPR-ed residual visibility phases for a subset of source and frequency pairs from the comprehensive analysis presented in this paper. 
The target source and target frequency band (i.e. $\nu^{\rm high}$), along with the angular separation to the reference source (in all cases 1803+784), the reference frequency band (i.e. $\nu^{\rm low}$) and the frequency ratio R, in parenthesis, are specified for each plot. 
From left to right: 1928+738 at W-band ($6.8^o$ apart on the sky, K-band, R=4);
2007+777 at W-band (6.3$^o$ apart, Q-band, R=2); 1807+698 at D-band (8.6$^o$ apart, Q-band, R=3); 1842+681 at D-band
($10.7^o$ apart, Q-band, R=3). {\it Lower Row:} SFPR-ed maps resulting from Fourier transformation of the SFPR-ed visibility phases directly above. From left to right: 
1928+738 at 87\,GHz (W-band), 2007+777 at 87\,GHz; 1807+698 and 1842+681 at 130\,GHz (D-band). Peak fluxes are 2 Jy beam$^{-1}$, 266 mJy beam$^{-1}$, 415 mJy beam$^{-1}$ and 216 mJy beam$^{-1}$, respectively.
The contour levels in the maps start from 0.75\% of the corresponding peak fluxes, respectively, and doubling thereafter in all cases. Each map includes a negative contour level at the same percent level of the peak flux as the first positive one. The beam size is indicated at the bottom left of the image.}
\label{fig:sfpr_vis}
\end{figure}

The rapid changes in the observed interferometric phases introduced by the tropospheric propagation effects set a severe limit on the coherence time for integration of the signal in observations at high frequencies, and therefore the sensitivity of those observations.
A direct consequence of the effective tropospheric compensation achieved from simultaneous dual frequency observations is an increased coherence time and therefore sensitivity. This can be visually appreciated in the FPT and SFPR-ed phases shown in Figures \ref{fig:fpt_vis} and \ref{fig:sfpr_vis}, respectively.
Figure \ref{fig:fpt_vis} shows the FPT-ed calibrated visibility phases at $\nu^{\rm high}$ for the five frequency pairs ($\nu^{\rm low} \rightarrow \nu^{\rm high}$) with R integer, shown in separate plots, in our observations.
Note that in all cases, the calibration applied has been derived from a different frequency band and scaled with the corresponding factor R. In all cases, the compensation of the fast tropospheric fluctuations results in a much higher degree of coherence, compared to the raw output of the correlator.
Figure \ref{fig:sfpr_vis} shows the SFPR-ed visibility phases for a subset of frequency pairs and source pairs (using 1803+784 as reference) in these observations, which are representative of the final products of the SFPR analysis. It is immediately obvious that the remaining dispersive residual phase variation in the FPT-ed visibilities has been compensated for in the SFPR visibilities. 

\begin{figure}
\centering
\includegraphics[width=0.9\textwidth]{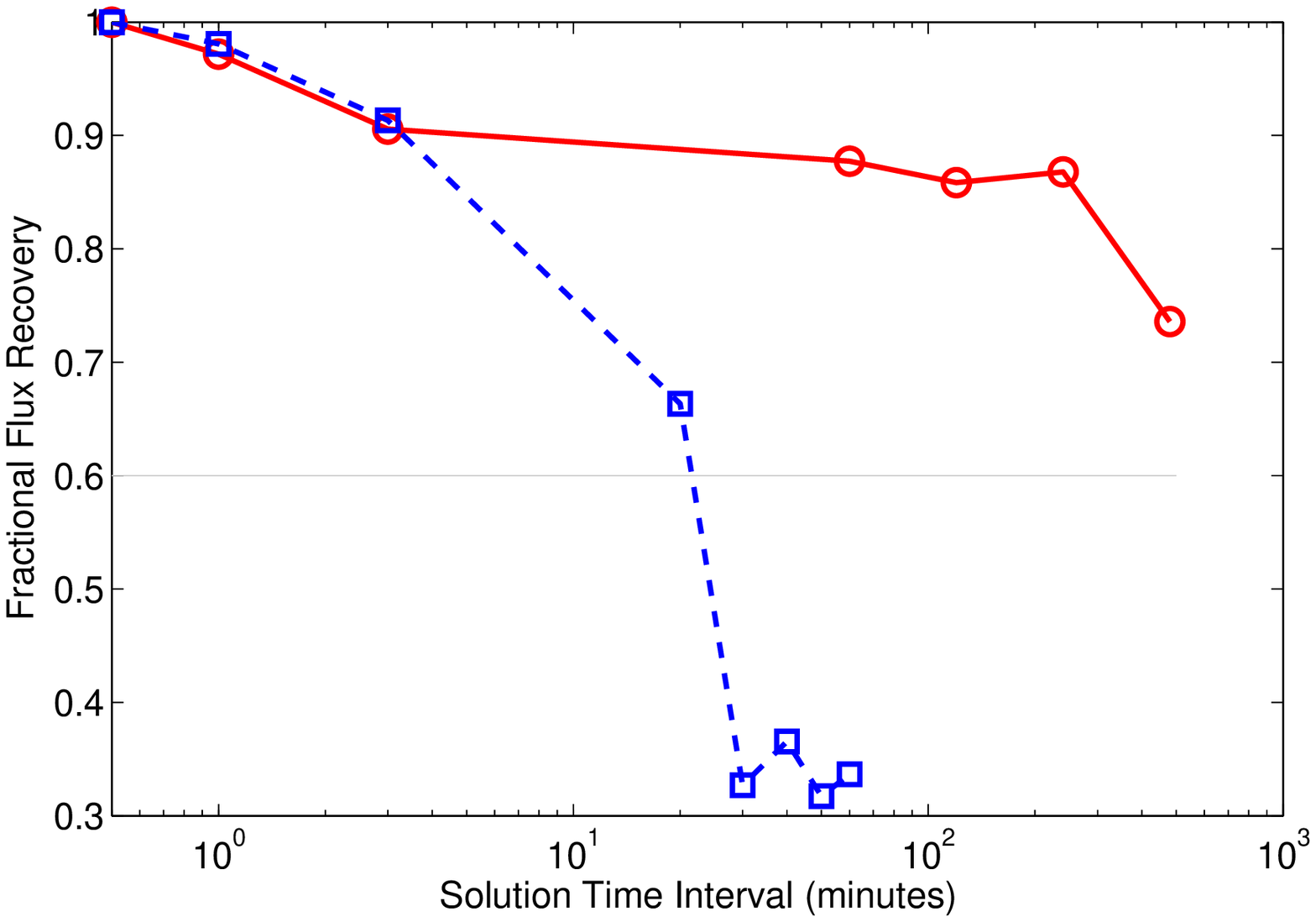}
\caption{
Plot of the Fractional Flux Recovery quantity (see text for details) versus length of (self-calibration) phase solution time interval, in minutes, for comparative 
coherence studies at 130\,GHz  between FPT (blue dashed line, with box symbols) and SFPR (red solid line, with circle symbols) calibration.
The coherence time corresponds to a  fractional flux recovery of $\sim$ 0.6.
For this study we used KVN observations of 1842+681 at $\nu^{\rm high} \,=\,130$\,GHz (D-band) calibrated with scaled up phase solutions at $\nu^{\rm low} \,=\,43$\,GHz (Q-band) for FPT, 
plus 1803+784 as the second source for SFPR analysis.}

\label{fig:coh}
\end{figure}

In order to quantify the increase in the coherence time, we have carried out a comparative study of the coherence times achieved with FPT and SFPR 
at 130\,GHz. This is the highest frequency in our observations, where the propagation effects are most severe.
To perform these tests we use the {\sc AIPS} task {\sc CALIB} on FPT-ed and SFPR-ed calibrated datasets of 1842+681 at $\nu^{\rm high} \, =\, 130$\,GHz, with $\nu^{\rm low} \, =\,43$\,GHz 
and 1803+784 as the reference source, 
with a series of phase solution time intervals ranging from 0.5 to 480 minutes. In each case the phase solutions are applied and the calibrated visibility data Fourier inverted to produce a map. 
We use the fractional peak recovered flux quantity, defined as the ratio of the peak flux in this map and that from self-calibrated maps, as a measure of remaining phase errors in the analysis.
Figure \ref{fig:coh} shows that the fractional peak recovered flux values in the maps 
decrease with increasing temporal solution intervals, as expected. 
The coherence time is defined as the solution interval at which the peak flux recovery is $60\%$, being equivalent to the rms residual phase being equal to 1 radian. 
Our analysis show that the coherence time at 130-GHz is $\sim$20 minutes with FPT calibration. With SFPR calibration there is practically no limit in the coherence time; we could integrate up to 8-hours, the whole duration of the experiment, with a mere $20\%$ loss of peak flux. Note that the tropospheric coherence time at 130-GHz is some tens of seconds and that neither 1842+681 nor 2007+777 have direct detections at this frequency.

\subsection{SFPR Maps and Astrometry at 22, 43, 87 and 130-GHz.} 

The final outcome of the SFPR analysis is a SFPR map which conveys the astrometric information.
Figure \ref{fig:sfpr_vis} shows a subset of the SFPR-ed maps obtained in the comprehensive analysis of KVN observations; they are the fourier transform of the SFPR visibilities directly above in the same figure.
The offset of the peak of brightness with respect to the center of the maps is a measurement of the relative position shift between the two frequency bands, for the two sources. 
The complete results from the SFPR astrometric analysis, comprising of the five frequency pairs and six source pairs are summarized in Table \ref{tab:pair}.
Table \ref{tab:pair} lists the right ascension and declination offsets of the peak of brightness from the center of the SFPR maps, as measured with {\sc AIPS} task {\sc MAXFIT}; the dynamic range of the map; the estimates of rms SFPR phase errors arising from the different contributions following formulae in \citet{rioja_11a}, along with their quadratic sum; the last columns are the estimated astrometric errors as described in Section 4.5.
%
Note that the dual frequency calibration provides a perfect compensation for 
the non-dispersive phase errors (i.e. $\phi^{\rm high}_{\rm geo}$, $\phi^{\rm high}_{\rm dtrp}$ and $\phi^{\rm high}_{\rm strp}$, which stand for geometric, dynamic and static tropospheric errors, respectively), that the thermal noise term ($\phi^{\rm high}_{\rm thm}$) and the dynamic ionospheric errors ($\phi^{\rm high}_{\rm dion}$) 
are rarely significant and that the phase errors are dominated by the contribution from the static ionosphere contribution ($\phi^{\rm high}_{\rm sion}$). The latter is largest for the frequency pairs with $\nu^{\rm low} \,=\, 22$\,GHz and increases with the angular separation between the sources. 

\begin{footnotesize}
\begin{table}
\textwidth=22cm
\centering
\vspace*{-4cm}
\hspace*{-2cm}
\begin{tabular}{lrrrrrrrrrrrr}%
\hline \hline
Freq.&\multicolumn{2}{c}{SFPR Astrometry}&&\multicolumn{7}{c}{rms SFPR  Phase Errors} &\multicolumn{2}{c}{SFPR Errors}\\
\cline{2-3}\cline{5-10}\cline{12-13}
Pair&$\Delta \alpha$\,cos$\delta$&$\Delta \delta$&DR&$\sigma {\phi^{\rm high}_{\rm thm}}$&$\sigma {\phi^{\rm high}_{\rm geo}}$&
$\sigma {\phi^{\rm high}_{\rm dtrp}}$ &$\sigma {\phi^{\rm high}_{\rm strp}}$ &
$\sigma {\phi^{\rm high}_{\rm dion}}$ &$\sigma {\phi^{\rm high}_{\rm
  sion}}$ &$\sqrt{\Sigma\sigma_\phi^2}$&$\sigma_{\Delta \alpha\,{\rm cos}\,\delta}$&$\sigma_{\Delta \delta}$\\
$\nu^{\rm low}\rightarrow\nu^{\rm high}$& ($\mu$as)&  ($\mu$as) & & ($^o$) &  ($^o$) &  ($^o$) &  ($^o$) &  ($^o$) &  ($^o$) &  ($^o$) &  ($\mu$as)&  ($\mu$as)\\
\hline
&\multicolumn{10}{c}{1803+784 /1807+698 (8.65$^o$ apart)}\\
K$\rightarrow$Q&86&7&390&1.6&0.0&0.0&0.0&4.1&24.6&25.0&88&127\\
K$\rightarrow$W&81&76&320&1.9&0.0&0.0&0.0&10.3&61.5&62.4&110&159\\
K$\rightarrow$D&129&132&23&27.1&0.0&0.0&0.0&16.1&95.7&100.7&118&171\\
Q$\rightarrow$W&-11&65&360&1.7&0.0&0.0&0.0&2.1&12.3&12.6&22&32\\
Q$\rightarrow$D&10&82&138&4.5&0.0&0.0&0.0&3.7&21.9&22.6&27&38\\
\hline
&\multicolumn{10}{c}{1803+784 / 1842+681 (10.70$^o$ apart)}\\
K$\rightarrow$Q&-200&75&360&1.7&0.0&0.0&0.0&4.4&30.4&30.8&109&157\\
K$\rightarrow$W&-364&169&200&3.1&0.0&0.0&0.0&11.1&76.1&76.9&136&196\\
K$\rightarrow$D&-379&241&40&15.6&0.0&0.0&0.0&17.2&118.3&120.6&142&205\\
Q$\rightarrow$W&-192&110&240&2.6&0.0&0.0&0.0&2.2&15.2&15.6&27&40\\
Q$\rightarrow$D&-235&130&57&10.9&0.0&0.0&0.0&3.9&27.0&29.4&35&50\\
\hline
&\multicolumn{10}{c}{1803+784 / 1928+738 (6.79$^o$ apart)}\\
K$\rightarrow$Q&-86&70&830&0.8&0.0&0.0&0.0&3.9&19.3&19.7&69&100\\
K$\rightarrow$W&-241&150&380&1.6&0.0&0.0&0.0&9.7&48.3&49.3&87&125\\
K$\rightarrow$D&-307&210&100&6.2&0.0&0.0&0.0&15.0&75.1&76.8&90&130\\
Q$\rightarrow$W&-156&64&960&0.6&0.0&0.0&0.0&1.9&9.7&9.9&17&25\\
Q$\rightarrow$D&-165&84&150&4.2&0.0&0.0&0.0&3.4&17.2&18.0&21&31\\
\hline
&\multicolumn{10}{c}{1803+784 / 2007+777 (6.34$^o$ apart)}\\
K$\rightarrow$Q&-43&43&390&1.6&0.0&0.0&0.0&3.8&18.0&18.5&65&94\\
K$\rightarrow$W&0&18&70&8.9&0.0&0.0&0.0&9.5&45.1&46.9&83&119\\
K$\rightarrow$D&-4&45&69&9.0&0.0&0.0&0.0&14.8&70.1&72.2&85&123\\
Q$\rightarrow$W&-41&15&540&1.2&0.0&0.0&0.0&1.9&9.0&9.3&16&24\\
Q$\rightarrow$D&-42&50&85&7.3&0.0&0.0&0.0&3.4&16.0&17.9&21&30\\
\hline
&\multicolumn{10}{c}{1807+698 / 1842+681 (3.60$^o$ apart)}\\
K$\rightarrow$Q&-276&73&376&1.7&0.0&0.0&0.0&3.4&10.2&10.9&38&56\\
K$\rightarrow$W&-476&123&227&2.7&0.0&0.0&0.0&8.5&25.6&27.1&48&69\\
K$\rightarrow$D&-549&156&67&9.3&0.0&0.0&0.0&13.2&39.8&43.0&50&73\\
Q$\rightarrow$W&-185&47&133&4.7&0.0&0.0&0.0&1.7&5.1&7.1&13&18\\
Q$\rightarrow$D&-230&45&143&4.4&0.0&0.0&0.0&3.0&9.1&10.5&12&18\\
\hline
&\multicolumn{10}{c}{1928+738 / 2007+777 (4.52$^o$ apart)}\\
K$\rightarrow$Q&125&-92&193&1.9&0.0&0.0&0.0&3.5&12.9&13.5&47&69\\
K$\rightarrow$W&181&-205&162&2.2&0.0&0.0&0.0&8.8&32.1&33.4&59&85\\
K$\rightarrow$D&192&-257&85&4.2&0.0&0.0&0.0&13.7&50.0&52.0&61&88\\
Q$\rightarrow$W&69&-76&73&4.9&0.0&0.0&0.0&1.8&6.4&8.3&15&21\\
Q$\rightarrow$D&210&-105&70&5.1&0.0&0.0&0.0&3.1&11.4&12.9&15&22\\
\hline
\end{tabular}
\caption{
{\footnotesize Summary of the measurements from the SFPR astrometric analysis presented in this paper, along with the error estimates, for the five frequency pairs (column 1) and for the six source pairs (separated by horizontal lines).
The relative astrometric offsets and the dynamic ranges (columns 2--3 and 4, respectively) are measured from the SFPR maps. A list of the estimated error contributions, per baseline, is provided:
the thermal errors (column 5) are estimated from the dynamic range
(see text); the geometric and propagation media contribution errors
(columns 6--10) are estimated using the formulae in \citet{rioja_11a},
for typical parameter uncertainties of the tropospheric zenith path
delay and the TEC equal to 3 cm and 3 TECU, respectively, source
angular separations as listed, simultaneous multi-band observing
($T^\nu_{swt}=0$) and source switching cycle of
$T_{swt}=450\,$seconds. Column 11 is the quadratic sum of the
forementioned errors. Columns 12 and 13 are the errors of the SFPR astrometric measurements, in right ascension and declination.}
\label{tab:pair}}
\end{table}

\begin{table}
\centering
\hspace*{-1cm}
\begin{tabular}{l|ccccc}
\hline 
Source Name&1803+784$^\dagger$ &1807+698$^\dagger$  &1842+681  &1928+738$^\dagger$ &2007+777\\
Jet PA (deg)&91.5$\pm$3.5&97.5$\pm$2.5&-135$\pm$5&-160$\pm$10&90$\pm$5\\
\hline
\end{tabular}
\caption{Jet position angle (PA) and error range from high resolution
  maps in \citet{hovatta_14} where marked
  with $^\dagger$; otherwise measured from the MOJAVE maps
  \citep{mojave} with an estimated error of 5$^o$. \label{tab:globc_a}}
\end{table}
\begin{table}
\centering
\begin{tabular}{l|ccccc}
\hline
 Freq. Pairs&K$\rightarrow$Q&K$\rightarrow$W&K$\rightarrow$D&
Q$\rightarrow$W&Q$\rightarrow$D\\
Global Correction ($\mu$as)&-42,-36&-82,-76&-88,-100&-48,-326&-82,-34\\
Errors  ($\mu$as)&$\pm$14,$\pm$4&$\pm$29,$\pm$10&$\pm$39,$\pm$13&$\pm$10,$\pm$6&$\pm$14,$\pm$9\\
\hline

\end{tabular}
\caption{
List of the global astrometric correction vectors, in $\mu$as on the
sky of right ascension and declination, that result in the best
alignment of the jet directions (listed in Table \ref{tab:globc_a})
and the individual source frequency dependent position shifts, for the
five frequency pairs.  The errors in these corrections are given below.
\label{tab:globc_b}}
\end{table}
\end{footnotesize}

\subsection{Decomposition of relative astrometric measurements into individual source  position-shifts} 

\begin{figure}
\centering
\includegraphics[width=0.8\textwidth]{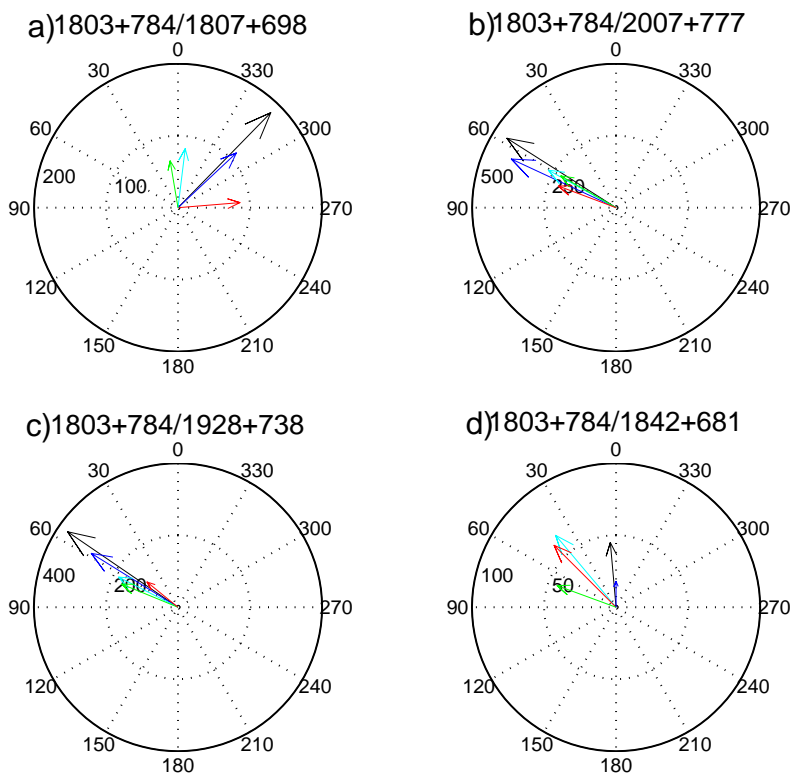}
\caption{Polar plots of the SFPR astrometric measurements in Table \ref{tab:pair},  corresponding to 
the four pairs of sources with 1803+784 (plots a-d, with 1807+698, 1842+681, 1928+738 and 2007+777, respectively)  at the five pairs of frequencies (shown in different colors):
K$\rightarrow$Q (red), K$\rightarrow$W (blue), K$\rightarrow$D (black), Q$\rightarrow$W (green) and Q$\rightarrow$D (cyan). 
The vectors correspond to the relative position shifts for the two sources between two frequencies, for the five frequency pairs. 
In the polar plots the position angles are shown outside the largest circle and are 0 and 90 degrees towards North and East, respectively, 
and the magnitude units, as specified in the concentric circles, are in $\mu$as.  
}
\label{fig:cs_pairs}
\end{figure}

Figure \ref{fig:cs_pairs} shows polar plots of the pairwise astrometric measurements listed in Table \ref{tab:pair}, for the four source pairs involving 1803+784, and for the five frequency pairs. These are direct outcomes of the SFPR analysis. 
For each source pair, the measurements are the combined position shift contributions from both sources between the two frequencies, for each frequency pair and therefore are expected to show little directional coherence, except when one source has a dominant position-shift  (e.g. the plots involving 1842+681 or 1928+738).  

\begin{figure}
\centering
\includegraphics[width=1.05\textwidth]{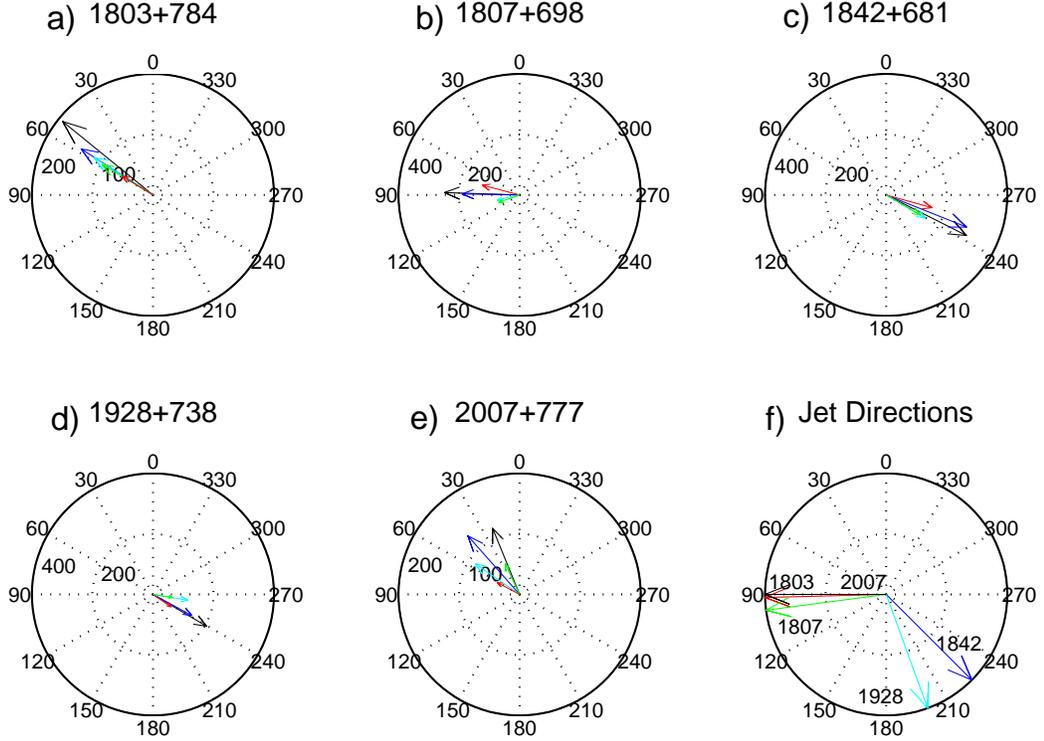}
\caption{(a-e) Polar plots of the degenerate decomposed single source position shifts between two frequencies, for the five frequency pairs and the five AGNs. These are
shown in different colors (as in Figure \ref{fig:cs_pairs}):  K$\rightarrow$Q (red), K$\rightarrow$W (blue), K$\rightarrow$D (black), Q$\rightarrow$W (green) and Q$\rightarrow$D (cyan).
They have been derived from the SFPR pair-wise measurements using SVD. 
Each vector represents the best linear least squares minimised position shift, and includes a global (i.e. common for all sources) degenerate offset for each frequency band. 
The axes of the polar plots are as in Figure \ref{fig:cs_pairs}.  
(f) Polar plot of the jet directions as appear in high resolution maps of the five AGNs, showing  the poor alignment with the position shifts in (a-e). 
The vector colours for the jet directions are: red for 1803+784, green for 1807+698, orange for 1842+681, blue for 1928+738 and cyan for 2007+777.}
\label{fig:cs_indi}
\end{figure}

\begin{figure}
\centering
\includegraphics[width=0.9\textwidth]{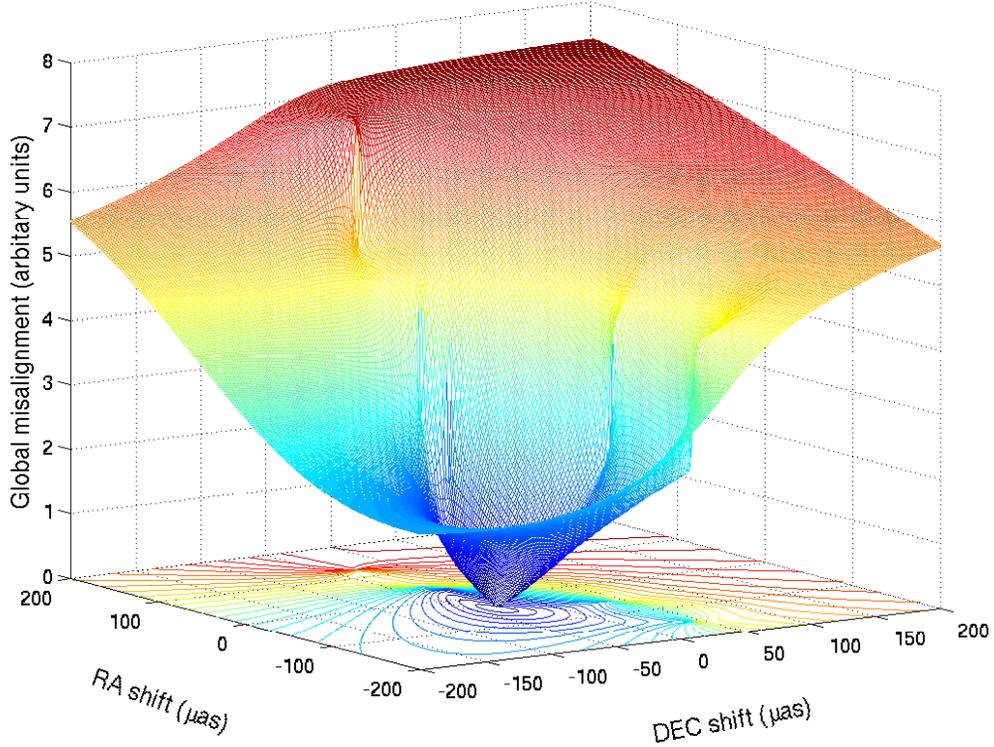}
\caption{
The combined misalignment for all sources (Z-axis) between the jet directions in Table 3 and the 
frequency dependent position shift vectors of the same source, evaluated on a grid of global (i.e. common for all sources) corrections to the degenerate SVD position shift vectors shown in Figure 8, for the frequency pair 22/43 GHz (K$\rightarrow$Q). The grid ranges from $\pm \, 200\, \mu$as in right ascension and declination (X and Y-axis, respectively). 
The global misalignment is shown in arbitrary units of the absolute sum of the differences between the complex direction vectors of jets and modified position shift measurements for all sources.
The best global alignment between the directions of the jet and the position shift corresponds to a correction 
of -42 and -36\,$\mu$as, in right ascension and declination, respectively, to the SVD solution. 
A similar procedure was carried out for each frequency pair; the results of the complete analysis are listed in Table \ref{tab:globc_b}. }
\label{fig:cs_grid}
\end{figure}

\begin{figure}
\centering
\includegraphics[width=1.05\textwidth]{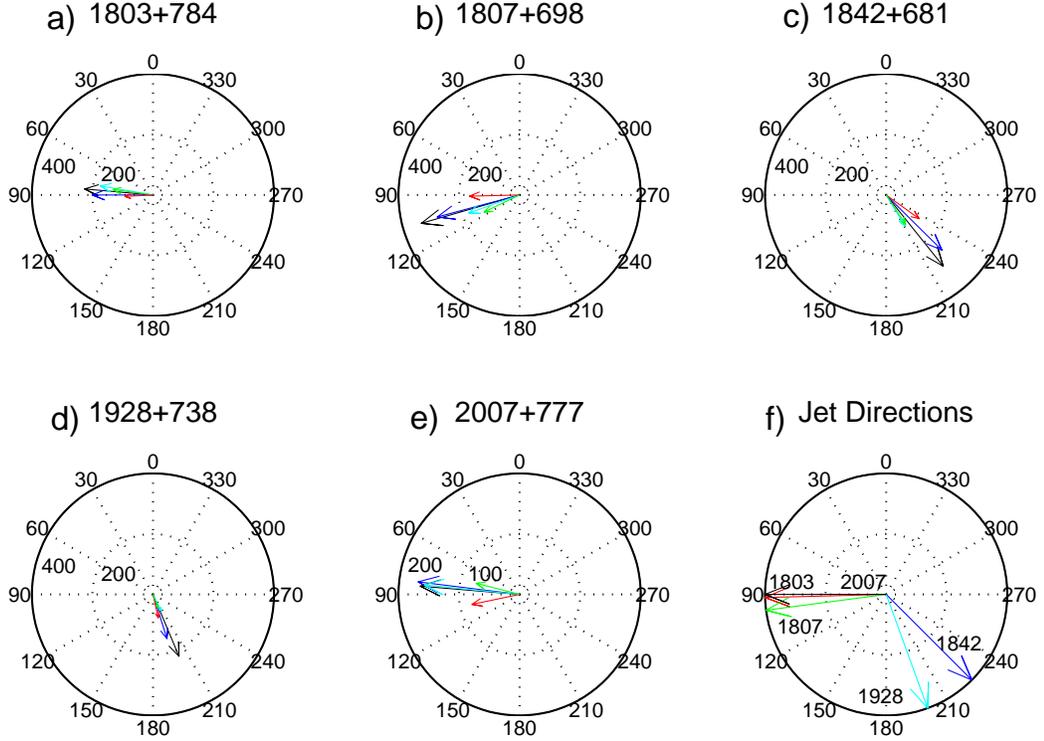}
\caption{(a-e) Polar plots of the decomposed absolute single source position shifts between two frequencies, for the five frequency pairs and the five AGNs. These are
shown in different colors (as in Figure \ref{fig:cs_pairs}):  K$\rightarrow$Q (red), K$\rightarrow$W (blue), K$\rightarrow$D (black), Q$\rightarrow$W (green) and Q$\rightarrow$D (cyan).
They have been derived from the SFPR pairwise measurements, using SVD plus the alignment constraint between the jet and the position shift directions, to break the degeneracy. 
The axes of the polar plots are as in Figure \ref{fig:cs_pairs}.  
(f) Polar plot of the jet directions as for Figure \ref{fig:cs_indi}, 
showing the  good alignment between the jet directions and that of the decomposed frequency dependent position-shifts, for each source.
Thus we have obtained absolute position shifts for the individual sources from the original pairwise measurements.}
\label{fig:cs_indi2}
\end{figure}

We have used SVD to decompose the pairwise position shifts into single source frequency dependent position shifts, 
albeit with degeneracies included. 
Those are shown in Figure \ref{fig:cs_indi},  where one can appreciate 
an improved agreement between the directions of the position-shifts 
for each source, 
although those are not well aligned with the jet direction in high resolution maps. The jet directions in the high resolution maps for the five AGN sources \citep{hovatta_14,mojave} are also shown in Figure  \ref{fig:cs_indi}e and their values are listed in Table \ref{tab:globc_a}. 
We used the expectation of alignment to break the degeneracy, 
%
by finding the best global correction (i.e. common for all sources) through a grid search of a few hundred micro-arcseconds around the SVD solutions, for each frequency pair. Figure \ref{fig:cs_grid} shows the degree of  alignment
as a function of grid position for the K$\rightarrow$Q dataset. 
Table \ref{tab:globc_b} lists the global corrections that were found to best align the SVD minimized single source position shifts to the jet directions, for the five frequency pairs. 
Finally, Table \ref{tab:indi} lists the resultant absolute single source position shifts corresponding to the five frequency pairs, for the five sources. 
Figure \ref{fig:cs_indi2} shows polar plots of these single source frequency dependent absolute position shifts listed in Table \ref{tab:indi}, which display a tight agreement between the five frequency pairs and are well aligned with the jet directions for each source. 

\begin{footnotesize}
\begin{table}
\centering
\begin{tabular}{lcccrc}
\hline \hline
1&2&3&4&5&6\\
&\multicolumn{3}{c|}{Absolute Position
  Shift}&\multicolumn{2}{|c}{Registration Error}\\ 
\cline{2-6}
Freq.&$\Delta \alpha$\,cos$\delta$&$\Delta\delta$&PA&$\sigma_{\Delta \alpha cos\delta}$&$\sigma_{\Delta \delta}$\\
Pair& ($\mu$as)& ($\mu$as) & (deg) & ($\mu$as)& ($\mu$as) \\
\hline
&\multicolumn{5}{c}{1803+784}\\
K$\rightarrow$Q&-92&-4&92&31&41\\
K$\rightarrow$W&-200&0&89&43&51\\
K$\rightarrow$D&-225&19&84&50&54\\
Q$\rightarrow$W&-132&18&82&11&12\\
Q$\rightarrow$D&-173&30&80&15&15\\
\hline
&\multicolumn{5}{c}{1807+698}\\
K$\rightarrow$Q&-164&-3&91&24&29\\
K$\rightarrow$W&-272&-72&104&36&37\\
K$\rightarrow$D&-324&-93&106&43&39\\
Q$\rightarrow$W&-116&-56&115&10&9\\
Q$\rightarrow$D&-167&-59&109&14&12\\
\hline
&\multicolumn{5}{c}{1842+681}\\
K$\rightarrow$Q&109&-78&-125&66&93\\
K$\rightarrow$W&183&-182&-134&85&117\\
K$\rightarrow$D&188&-235&-141&91&122\\
Q$\rightarrow$W&64&-97&-146&17&23\\
Q$\rightarrow$D&61&-102&-148&24&30\\
\hline
&\multicolumn{5}{c}{1928+738}\\
K$\rightarrow$Q&17&-77&-167&32&43\\
K$\rightarrow$W&46&-145&-162&45&54\\
K$\rightarrow$D&85&-204&-157&52&57\\
Q$\rightarrow$W&16&-42&-158&12&12\\
Q$\rightarrow$D&27&-54&-153&15&16\\
\hline
&\multicolumn{5}{c}{2007+777}\\
K$\rightarrow$Q&-78&-16&101&18&20\\
K$\rightarrow$W&-167&20&82&31&26\\
K$\rightarrow$D&-163&13&85&39&28\\
Q$\rightarrow$W&-71&18&75&10&9\\
Q$\rightarrow$D&-157&15&84&13&10\\
\hline
\end{tabular}
\caption{\small
List of absolute single source position shifts between two
frequencies, for the five frequency pairs and for the five AGNs in this
study, as plotted in Figure \ref{fig:cs_indi2}.
They have been derived from the SFPR pairwise measurements, using SVD
plus the alignment constraint between the jet and the position shift
directions, to decompose into single source contributions and to break
the degeneracy. 
Column 1 are the frequency pairs; Columns 2 and 3 are the Right
Ascension and Declination of the position shifts,
respectively, with the
corresponding position shift direction (PA) in Column 4. The errors
(Columns 5 and 6) include all random and systematic contributions. 
The position shift direction (PA) should be compared with the jet
position angles given in Table \ref{tab:globc_a}. 
\label{tab:indi}}
\end{table}
\end{footnotesize}

\subsection{Astrometrical Error Analysis}

We have carried out a comprehensive error analysis to estimate the propagation of random and systematic error contributions in the SFPR analysis, along with those from the SVD and global-shift minimization analysis, into the frequency dependent position shift astrometric measurements for each source.

For the SFPR error analysis, we have used the formulae in \citet{rioja_11a} to estimate the residual phase errors arising  from typical parameter uncertainties in the `a priori' models for the propagation medium and the geometry contributions.
The estimated values per baseline are listed in Table \ref{tab:pair}, for the geometry ($\sigma {\phi^{high}_{geo}}$) and for the dynamic and static components 
of both the troposphere ($\sigma {\phi^{high}_{dtrp}}$ and $\sigma {\phi^{high}_{strp}}$, respectively)  and the ionosphere ($\sigma {\phi^{high}_{dion}}$ and $\sigma {\phi^{high}_{sion}}$, respectively). It should be noted that the table entries corresponding to non-dispersive errors are zero, as a result of the multi-frequency calibration. 
Table \ref{tab:pair} includes also the dynamic range (DR) values measured from the SFPR maps, which are used to estimate a per baseline thermal phase error ($\sigma {\phi^{high}_{thm}}$)
using the expression $360^o$/DR/$\sqrt{{\rm N}_{\rm ant}}$, where N$_{\rm ant}$ is the number of antennas.
This is derived using the relationship between positional error and dynamic range ($\sigma_{\alpha,\delta} \sim \theta_{\rm beam}$/DR) and the formulae in \citet[A12.58]{tms}.
%
Note that the dominant error contribution in Table \ref{tab:pair} is related to the static component of the ionospheric propagation, which reaches peak values for frequency pairs with $\nu^{low}=22\,$GHz  and larger values of R and source pair angular separations; this will be revisited in the discussions section.
The quadratic sum 
of the forementioned errors ($\sqrt{\Sigma\sigma^2}$) 
is converted to the final SFPR astrometric error ($\sigma_{\Delta \alpha\, {\rm cos}\,\delta}, \sigma_{\Delta \delta}$), for the KVN baselines lengths of ca. 400 km.



Finally, we convert the SFPR astrometric errors to frequency dependent position shift errors for each source by: 
1) passing those through the same SVD transformation used for the decomposition of the measurements, 
and 2) combine the outcome with the errors in the global shift minimization analysis, i.e. the errors in the  jet position angles measured from the maps, as listed in Table \ref{tab:globc_a}.
%
The final astrometric accuracies are listed in Table \ref{tab:indi}, in $\mu$as, as $\sigma_{{\Delta \alpha} \, {\rm cos}\, \delta}$ and $\sigma_{\Delta \delta}$. \\
These correspond to the errors in the measurements of the position shifts that enable the bona fide astrometric registration of the maps across frequencies, for each of the five observed sources.

\section{Discussions}\label{sec:disc}

\subsection{Demonstration of Multi-Frequency Calibration and Astrometry up to 130\,GHz}

Simultaneous multi-frequency observations offer an effective path to achieve increased sensitivity and precision astrometry in mm-VLBI,  
beyond the domain of standard techniques, such as phase referencing.  
The compensation of the fast phase changes imposed by the rapid tropospheric fluctuations in mm-VLBI, using observations at a lower frequency of the same source, results in an increased coherence time of up to 20 minutes at 130\,GHz using FPT analysis, which results in a significant increase of sensitivity. Moreover, when combined with the observations of another source, a bona fide astrometric measurement of the relative frequency dependent position-shift between the two frequencies 
can be estimated using the SFPR technique. This in turn results in an unlimited extension of the coherence time. 

The work presented in this paper corresponds to a first demonstration of SFPR at 130\,GHz, the highest frequency of the KVN. 
The application of SFPR techniques has allowed the detection of weak sources that were not directly detected within the atmospheric coherence time (i.e. with self calibration) and we have measured  frequency dependent position shifts between a range of frequencies from 22 up to 130\,GHz with high precision, for each of the observed AGNs.
Previous attempts to carry out astrometry at such frequencies, with the VLBA up to 87\, GHz, were very limited:
Once with conventional phase referencing using very fast source switching and a very close source pair with $\sim$14$^\prime$ separation \citep{porcas_02} and
once with fast frequency switching using SFPR \citep{rioja_11a}. Moreover, the fast frequency switching observing mode of the VLBA leads to residuals in the tropospheric compensation, which ultimately limit the accuracy and quality of the analysis.
The simultaneous multi-frequency observing capability simplifies and widens the application to even higher frequencies and to many targets. 
Therefore we have demonstrated the benefits of simultaneous multi-frequency observations for mm-VLBI, to achieve sensitivity and astrometry at frequencies up to 130\,GHz, the maximum frequency available with KVN.  We believe these benefits would continue to apply beyond this frequency.

\subsection{Interpretation of the Measured Position Shifts}


The outcome of the astrometric analysis presented in this paper is a measurement of the frequency dependent absolute position shift of the brightest features, or reference points, in the maps at the four KVN frequency bands, for each of the five AGNs. 
In general, the reference points in the KVN maps do not correspond to the position of the ``core''
component, due to structure blending effects resulting from the relatively short $\sim$400 km baselines. Therefore, in general, our measurements correspond to the position shifts or angular separations between the centroids of the brightness distributions at the four frequencies; in the case of a point source, this would be the same as the ``core-shift''.  

Higher resolution observations with longer baselines would make it possible to isolate the ``core'' component as the reference point in the astrometric analysis, and to achieve an increase in the  astrometric precision directly proportional to the enlarged baseline. For example, a 8000 km baseline would result in a twentyfold decrease of the astrometric errors listed in Tables \ref{tab:pair} and  \ref{tab:indi}, which would be suitable to measure the small magnitude of the opacity ``core-shift'' effect between the four KVN frequency bands predicted in the standard model for extragalactic radio sources \citep{pilot_proj}.
Regardless of the baseline length, the SFPR measurements provide a bona fide astrometric registration of the maps, which is at the base for reliable spectral index maps and in general spectral distribution studies. 
Applying the absolute single source postition shifts in Table \ref{tab:indi} to the hybrid maps in Figure \ref{fig:sc} provides the required astrometric image registration for such analysis.
%

In this paper, despite having the measurements of the shifts for the ``bona fide'' astrometric registration of the maps,   the poor amplitude calibration in our observations prevented us from obtaining meaningful spectral index maps.  In a second epoch of observations we have included an improved amplitude calibration strategy to overcome this issue.  

The canonical SFPR astrometric errors range from a few tens to a few hundreds of $\mu$as depending on the frequency pairs, and are completely dominated by the systematic  static ionospheric  terms ($\sigma \phi_{\rm sion}$). If this was actually the case we would expect to find (i) very similar SFPR astrometric offsets in all source pairs with similar separations, for a given frequency pair (i.e. 1928+738 and 2007+777 to 1803+784) and (ii) predictable ratios between the astrometric offsets, e.g. the ratio of the astrometric offsets for K$\rightarrow$W and Q$\rightarrow$W should be equal to 5 if the ionospheric uncertainties were dominant.
As we see no indication of these signatures in our results we believe that the ionospheric contributions are acting coherently across the sources and that the canonical errors are therefore overestimates. 
This is quite possible as the KVN baseline lengths are of the same order as the height of the ionospheric E-layer, therefore the atmospheres over the different antenna sites would not be fully decorrelated. 
For this reason, we used an uncertainty in the total electron content (TEC) parameter in our error analysis equal to 3 TEC units (TECU), which we believe is still an over-estimate.

In our results we have presented an experimental demonstration of successful ``bona fide'' astrometric registration between mm-VLBI maps, at the four KVN frequency bands, including, for the first time, high precision astrometric measurements at 130\,GHz.
The combination with similar studies at lower frequencies (see \citet{ros_01, guirado_00} for wide-field high precision astrometry of some of these sources) opens a promising path towards providing a complete picture of the underlying physical mechanisms of jet formation across a very wide frequency range. 



\subsection{Comparison of SFPR to other Methods}

There exist a variety of methods used to register the images of AGNs and/or maser species at different frequencies; here we discuss those, in comparison to SFPR, for mm-VLBI.
%
\citet{hovatta_14} provides a thorough review of the alternative methods for registration of the images of AGNs at multiple frequencies. These broadly consists of: using optically thin bright jet components \citep{fromm_13}, 2D cross-correlation algorithms \citep{croke_08}, or a combination of both. However these methods are predicated on the assumption that there is a clearly identifiable optically thin bright jet component, which can act as a reference point for all frequencies, or an ensemble of less bright optically thin jet components, which can provide an average registration. 
Hovatta \etal carried out an error analysis of the propagation of incorrect alignments on spectral index and rotation measure maps. They concluded that the alignment errors are dominant around the core region (up to a distance of $\sim$3\,mas) and therefore the conclusions on the spectral index distribution for the innermost jet regions should be treated with caution. 
The application of these methods for compact sources and for sources with faint or smooth jets is clearly an issue. Therefore these methods can be unreliable or impossible to use in mm-VLBI, where in many cases only the compact core can be detected. 

In a similar fashion, 
some maser species can be assumed to form in a ring, and the centre of the ring can therefore act as a reference point \citep{desmurs_00}, or one can identify a single component which appears similar in velocity and orientation with respect to the main body of emission and use that as the reference point across frequencies \citep{cotton_04}. It is not hard to see the short-falls in such approaches and these different methods tend to produce incompatible conclusions. 
Phase referencing would provide a clear solution for such challenges but, as pointed out previously, phase-referencing is not an option above 43\,GHz, in general.

Therefore SFPR stands alone as a method which will allow for the unambiguous registration across wide frequency spans for mm-VLBI images, both for continuum and spectral line studies, because it provides a complete compensation of atmospheric propagation and instrumental effects. 
%
SFPR is widely applicable for many sources, since the calibrator source can be at a significant angular separation and slow source switching does not undermine the result. The method will work even at very high frequencies, making it particularly suitable for mm-VLBI observations. We do note that systematic effects do need to be carefully taken into account, particularly when using a lower frequency of 22\,GHz. 

\section{Conclusions}\label{sec:conc}

We have demonstrated that the KVN multi-band system is capable of delivering increased coherence times by calibrating the highest frequencies with the scaled up phase solutions from the lower frequencies, using the Frequency Phase Transfer method.
At  130\,GHz the coherence times were extended from a few tens of seconds to 20 minutes, and to many hours by interleaving observations of a second source.
This provides improved sensitivity through allowing longer integrations on weak sources. 

We have demonstrated that the KVN multi-band system is capable of delivering astrometric results at the highest frequencies, using the Source Frequency Phase Referencing technique. We have measured accurate relative position shifts between frequencies in the range of 22 to 130\,GHz, using observations of six pairs of sources with angular separations between 3.6$^o$ and 11$^o$.

We have shown how to decompose these relative measurements into absolute single source frequency dependent position-shifts. 
These decomposed position-shift measurements are all that is required to form high fidelity spectral index maps between the four frequency bands, which 
we will present in a subsequent paper. 

\noindent
{\bf Acknowledgements}

\noindent
We are grateful to all staff members and students in the KVN who
helped to operate the array.  The KVN is a facility operated by the
Korea Astronomy and Space Science Institute.  We acknowledge the
support of the Australian DFAT Grant AKF-201400010.


\end{document}